# Expanding the Pressure Frontier in Grüneisen Parameter Measurement: Study of Sodium Chloride

Jun Kong[1,2], Kaiyuan Shi[2], Xingbang Dong[2], Xiao Dong[1,*], Xin Zhang[2],

Jiaqing Zhang[2], Lei Su[2,3,†] and Guoqiang Yang[3,‡]

[1]*Key Laboratory of Weak-Light Nonlinear Photonics and School of Physics, Nankai University, Tianjin 300071, China*

[2]*Center for High Pressure Science and Technology Advanced Research, Beijing 100093, China*

[3]*Key Laboratory of Photochemistry, Institute of Chemistry, University of Chinese Academy of Sciences, Chinese Academy of Sciences, Beijing 100190, China*

The Grüneisen parameter ($\gamma$) is crucial for determining many thermal properties, including the anharmonic effect, thermostatistics, and equation of state (EOS) of materials. However, the isentropic adiabatic compression conditions required to measure the Grüneisen parameter under high pressure are difficult to achieve. Thus, direct experimental Grüneisen parameter data in a wide range of pressures is sparse. In this work, we developed a new device that can apply pressure (up to tens of GPa) with an extremely short time about 0.5 ms, confidently achieving isentropic adiabatic compression. Then, we applied our new technique to sodium chloride and measured its Grüneisen parameter, which conforms to previous theoretical predictions. According to our obtained sodium chloride Grüneisen parameters, the calculated Hugoniot curve of the NaCl B1 phase appears up to 20 GPa and 960 K, which compares very well with the shock compression experiment data by Fritz et al. and other calculation works. Our results suggest that this new method can reliably measure the Grüneisen parameter of even more materials, which is significant for researching the equation of state in substances.

The Grüneisen parameter describes the effect that changing the volume of a crystal lattice has on its vibrational properties and the effect of changing the temperature on the crystal lattice's size or dynamics. It is a dimensionless parameter as a scale of anharmonicity in phonons, with great importance in the Mie-Grüneisen equation of state. It is related to the lattice-specific heat, which bridges lattice mechanics and thermodynamics. Investigating the Grüneisen parameter is also critical in geophysical models, such as measuring the sound velocity and melting of solids.

Mie and Grüneisen developed a theory for solids one century ago, indicating that pressure might be considered a linear function of internal energy [1]. They discovered that a solid's internal energy $E$, is the sum of the thermal vibrational energy $E_T(V,T)$ and the cold energy $E_c(V)$, which results from atomic interactions while at rest. The related pressure $P$ is also the sum of two such terms, $P_T(V,T)$ and $P_c(V)$. The Mie–Grüneisen equation of state can be expressed by:

$$P(V,E) = P_c(V) + \frac{\gamma(V)}{V}(E - E_c(V)) \qquad (1)$$

and the Grüneisen parameter can be given by:

$$\gamma = V \left(\frac{\partial P}{\partial E}\right)_V \qquad (2)$$

which is the basic definition of the Grüneisen parameter. After derivation, the Grüneisen parameter can be given by:

$$\gamma = \left(\frac{K_S}{T}\right)\left(\frac{\partial T}{\partial P}\right)_S \qquad (3)$$

where $K_S$ is the adiabatic bulk modulus. Therefore, by measuring a change in sample temperature combined with a tiny adiabatic isentropic pressure change in experiments, we can determine the Grüneisen parameter of a material. In this work, we determined the pressure dependence of sodium chloride's Grüneisen parameter by measuring the $\Delta T/\Delta P$ for small pressure changes with a very short time of about 0.5 ms. We used Eq. (3), taking $\Delta T/\Delta P$ as $(\partial T/\partial P)_S$. giving us a method of determining the Grüneisen parameter, because $(\partial T/\partial P)_S$ can be measured directly and reliable data for $K_S$ is available for many materials.

To date, many theoretical calculations and experimental works have predicted and measured the Grüneisen parameters of materials, but there are still many difficulties to overcome theoretically and experimentally. Determining the phonon spectrums is essential for theoretical Grüneisen parameters calculations. However, they are often unstable under high pressure, making it tough to calculate the phonon spectrum of complex materials described by large supercells. In experiments, the pressure dependence of the Grüneisen parameter can be obtained by various methods, including calculations measuring the *P-V* data or other approaches based on thermodynamic formulations, which require accurate thermal expansion, bulk moduli, and heat capacity measurements. This thermodynamic data can only be available for many materials in the lower pressure range in the vast majority of cases. Unfortunately, most of these findings about Grüneisen parameter, even for a simple material like sodium chloride, are in sharp disagreement with one another [2]. In 1977, Boehler developed a new method of directly determining the Grüneisen parameter using Eq. (3). Subsequent work has improved this method greatly in pressure range, adiabatic condition and temperature correction [3]. Nevertheless, due to experimental setup limitations, the pressure of these experiments is limited to 3 GPa, and achieving higher pressure for adiabatic and isentropic compression proves difficult. Due to these challenges, we still lack direct experimental data under a wide range of pressures, which hinders studying the Mie–Grüneisen equation of state. Therefore, developing a reliable experimental approach for accurately measuring the Grüneisen parameter in a wide range of pressures is crucial.

Sodium chloride crystal is essential in high-pressure physics studies due to its simple structure and high compressibility. One study shows that the phase transition of sodium chloride from a NaCl structure to a CsCl structure (B1-B2 phase transition) can occur at 30 GPa and 300 K [4]. In fact, numerous experiments and theoretical calculations have studied the Grüneisen parameter of sodium chloride. First, Boehler [2] studied it using rapid compression below 3.3 GPa, before extending the pressure and temperature to 5 GPa and 800 °C [5]. The Grüneisen parameter of molten sodium chloride was also investigated from 70-80 GPa [6]. Birch calculated the Grüneisen parameter of sodium chloride below 30 GPa by analyzing thermodynamic data [7]. Boness measured the volumetric sound velocity of liquid single-crystal NaCl based on optical analysis technology and calculated the Grüneisen parameter of NaCl under high pressure [8]. So far, the Grüneisen parameter of sodium chloride has only been measured in small pressure interval ranges in experiments, so we do not have a comprehensive view of how it changes with pressure. Therefore, exploring the Grüneisen parameter of NaCl in a wide pressure range is very interesting and significant for establishing NaCl's broader equation of state.

We measured the pressure-dependent Grüneisen parameter of sodium chloride across a wide range of pressures from 3.8 to 21.33 GPa using a new method showcased in this work. The calculated Hugoniot curve using sodium chloride's Grüneisen parameters agrees well with shock compression experiment data and other calculations [9-11]. Our new and improved method combines the dynamic DAC (dDAC) [12] with an ultra-fast temperature measurement system to achieve an automatic, programmable high-pressure jump with simultaneous high-speed temperature detection. Furthermore, this method extends the pressure range of our Grüneisen parameter research, obtaining higher pressure measurements under adiabatic and isentropic compression than before.

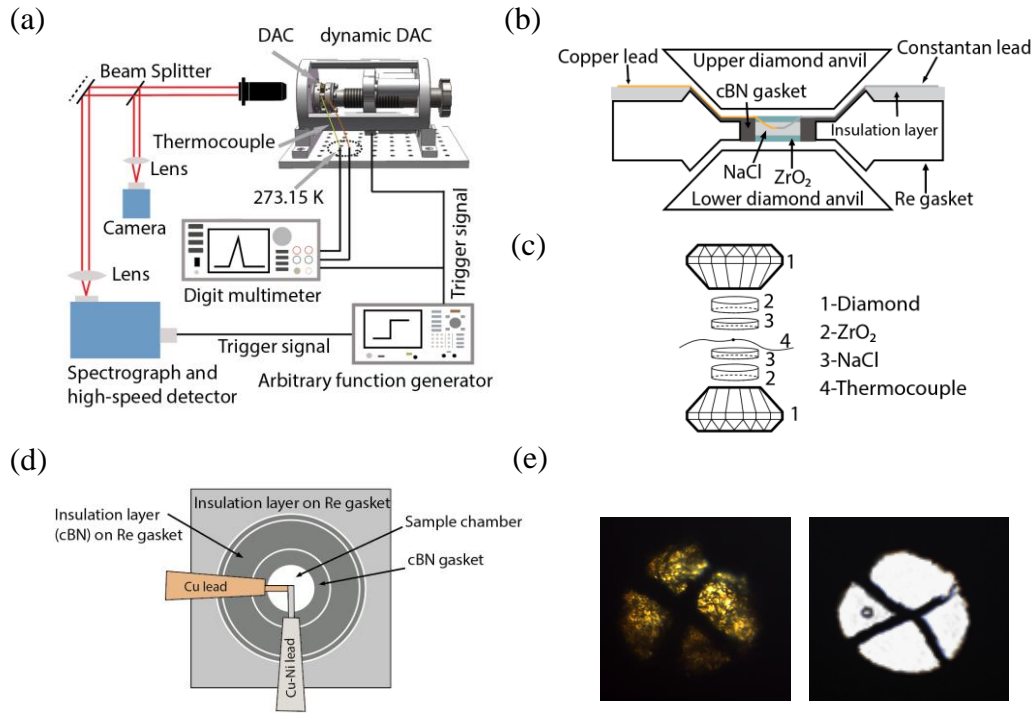

FIG. 1. (a) Schematic illustration of the programmable automatic high pressure jump and temperature detection apparatus. (b) Schematic illustration of the sample assembly in a DAC. (c) A side view of the sample assembly in the DAC. (d) A top view of the sample assembly in the DAC. (e) Microscopic images of the sample and thermocouple in the DAC before (left) and after (right) pressurization.

A developed dDAC was used to realize a pressure jump [12]. Time-resolved measurements, including thermal electromotive force (EMF) and fluorescent spectroscopy, were conducted to study the Grüneisen parameter of sodium chloride. Fig. 1a illustrates a schematic of this combined system, and more details can be found in the Supplementary Material [18]. Diamond anvil cells (DAC) with specific anvil culets were used for distinct pressure ranges. The sodium chloride powder sample and zirconia powder (used to prevent heat loss during the experiment) were loaded into the sample chamber in a Re gasket with an indented thickness of about 50 μm with a sandwich configuration (Fig. 1b and Fig. 1c). Fig. 1d shows the top view of the sample assembly in the DAC. A thin thermocouple with a thickness of 5 μm was inserted into the sodium chloride sample. The thermocouple was made by ourselves. Copper thin foil and constantan thin foil (Innochem, 99.99%, 0.005 mm) were cut with a width within 10 μm and touched together (Fig. 1d). We insulated the gasket using cubic boron nitride so that the thermocouple and gasket did not come into direct contact. In all experiments, the sample was first pre-pressed at a lower pressure to eliminate fine gaps and cavities. Fig. 1e shows the view through the gasket hole before and after compression. One ruby ball was buried in sodium chloride to gauge the pressure. A laser beam of 532 nm was focused on the ruby ball to stimulate the ruby fluorescence. Next, the signal was focused onto the spectrograph by a lens. Lamp lights illuminate the sample from behind and through the gasket hole to observe the morphology. Then the outgoing lights were reflected by a 50/50 beam splitter and finally collected by a camera. The temperature of the sample during compression can be obtained from the EMF data recorded by a digital multimeter. A freezing point bath was performed during the EMF measurement (Fig. 1a).

Due to its strong covalent bonds and low phonon dispersion, diamond is an excellent heat conductor, unlike most electrical insulators. Natural diamond has a thermal conductivity of roughly 2200 W/(m·K), five times greater than silver, the metal with the highest thermal conductivity [13]. Although we used zirconia with a thermal conductivity of 7.2 W/(m·K) [14] to prevent heat loss during the experiments, the heat was still lost quickly because diamond's thermal conductivity is far superior. Fig. 2a shows the variations of the thermocouple's EMF under different compression rates from about 2.5 GPa to 14 GPa. If the speed of the pressure jump is too slow, the heat will be lost too much, and the temperature measurement will be less accurate. For our devices, the pressurization time must be controlled within 1 ms to give the most accurate temperature measurement and best signal-to-noise ratio (SNR). The SNR is defined as the amplitude ratio of the EMF change under compression to the noise of the instruments. Therefore, in all our experiments, the pressure jump times were controlled within 1 ms. Consequently, our experimental conditions are close to adiabatic isentropic compression because of the pressure jump's high speed. Moreover, as shown in Fig. 2b, the pressure change and temperature change are detected nearly simultaneously in the experiment, so the response time of the thermocouple to temperature is negligible in our experiments. The thermocouple can reflect the sample's temperature changes with almost no delay. These features suggest that the entire signal of delta $T$ is from the sample rather than other parts of the cell arrangement during the pressure jump [15]. Moreover, the initial pressure and final pressure measurements were also taken separately by the ruby fluorescence spectrum before and after a pressure jump, and the pressure change with time can be seen as linear change (Fig. S1). Therefore, accurate pressure data can be provided to determine the Grüneisen parameters.

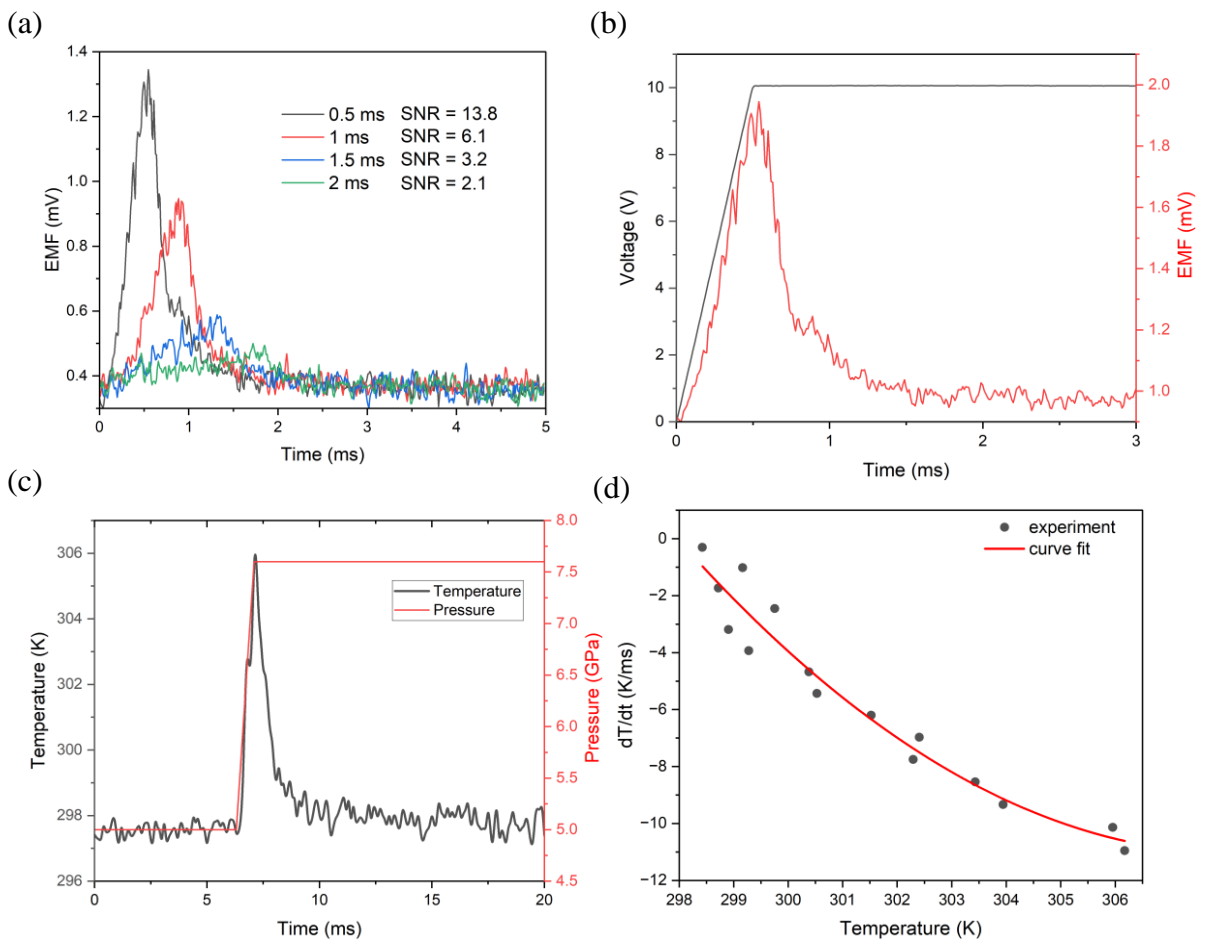

FIG. 2. (a) Variation of the thermocouple's thermal electromotive force (EMF) with time at different compression rates. (b) Synchronization test of the output voltage of the function generator for compression and EMF signal. The result shows that the temperature change is almost synchronized with the function generator's signal. (c) A typical record of temperature and pressure for sodium chloride before and after the pressure jump. (d) Temperature-decrease rate of sodium chloride with corresponding temperatures and its fitting curve. The curve indicates that the rate of temperature-decrease after the pressure jump. It was acquired by analysing the recorded data of Fig. 2c.

Fig. 2c shows the sample's temperature change with time in a typical run, where the pressure rises from 5 GPa to 7.6 GPa for 0.8 ms. The temperature decreased after the maximum while we maintained a stable pressure. The temperature change curve reveals the actual heat loss from the system, indicating that the process is not totally adiabatic and that temperature correction is required. We believe the temperature loss rate is identical between the heating and cooling processes because the geometric configuration of the experimental setup changes relatively little during the compression process [3]. Fig. 2d shows the analysed temperature decrease rate versus temperature after the pressure jump. On the basis of the data, we corrected the temperature change by integrating the rate with a selected tiny time interval during pressure jump and then added the integral absolute value to the recorded temperature change during the pressure jump at the time interval. Through the correction in temperature we can obtain adiabatic information from the actual process. The specific calculation process is described in detail in the Supplementary Material [18].

The Grüneisen parameter of sodium chloride at high pressure can be determined by Eq. (3). Here, $(\partial T/\partial P)_S$ can be approximated as $\Delta T/\Delta P$ for a small pressure change [2]. Therefore, the Grüneisen parameter can be expressed as a modified relationship:

$$\gamma(P_m) = \left(\frac{K_S(P_m)}{T(P_m)}\right)\left(\frac{\Delta T}{\Delta P}\right)_S \qquad (4)$$

where $P_m$ is the mean pressure of the final and initial pressures for the small pressure change, $K_S(P_m)$ is adiabatic bulk modulus at the mean pressure [7], and $T(P_m)$ is the instantaneous temperature corresponding to the mean pressure in the adiabatic compression. The temperature effect on $K_S$ can be ignored because $(\partial K_S/\partial T)_P \approx 0.01$ GPa/deg, which is very small compared with $(\partial K_S/\partial P)_T$ [7].

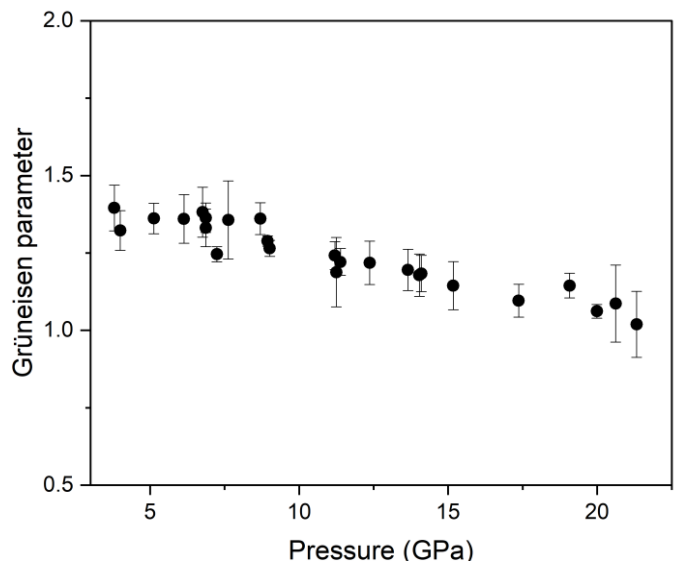

FIG. 3. The Grüneisen parameters of sodium chloride versus pressure.

The experimental data and results are listed in Table S1, and the obtained Grüneisen parameters of sodium chloride are plotted against pressure in Fig. 3. The temperature effect is disregarded because of the temperature's subtle influence of on the Grüneisen parameter [5]. The error bar stems from the inaccuracy of curve fitting about temperature versus time data.

We calculated it from Root Mean Square Error (RMSE) of curve fitting results. Then, we can obtain the error of $\gamma$ using $\gamma_{\pm}(P_m) = \left(\frac{K_S(P_m)}{T(P_m)}\right)\left(\frac{\Delta T \pm \sqrt{2}(RMSE)}{\Delta P}\right)_S$ for every point according to the method of error propagation.

To closely analyze the high-pressure evolution characteristics of the Grüneisen parameter, we compared our experimental results with previously published data in Fig. 4. Our comparison shows that we measured the Grüneisen parameters in a wider pressure range than previous experimental works. Our experiment conditions are relatively close to adiabatic isentropic compression because of the pressure jump's high speed. According to the dashed line in Fig. 4, the extrapolation from the previous experimental data does not obtain consistent results with our experimental data, but Birch's results from the EOS data derived from direct *P-V-T* measurements by Boehler & Kennedy [16] support our experimental results well. Birch obtained the Grüneisen parameter of sodium chloride in a wide pressure range from the EOS data according to $\gamma_{th} = (\beta B_T/c_V)V$, and the parameters contained in this equation can be measured independently. However, each physical quantity correlates with temperature and pressure, making it difficult to accurately measure for many other materials, especially the heat capacity at a constant volume ($c_V$). Therefore, our method is a better way to determine the Grüneisen parameter experimentally.

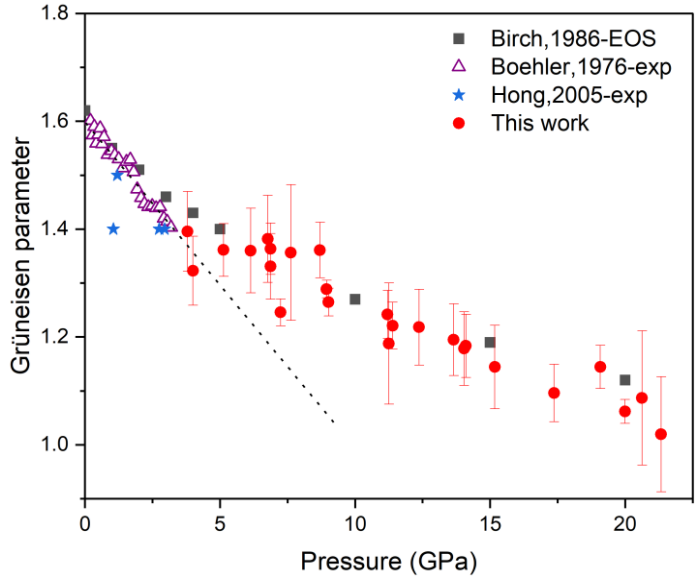

FIG. 4. Grüneisen parameter vs. pressure for sodium chloride from our data and previously published data. We measured the Grüneisen parameters in a wide pressure range. The dashed line is the extrapolation from the previous experimental data.

The shock metamorphism of sodium chloride provides crucial research information. The most important physical properties for studying this phenomenon are Hugoniot data using a flyer impact drive via a light gas gun or explosive loading. The linear relation between the shock wave and particle velocity and the Hugoniot curve has been widely used to solve various analytical equations of state in condensed matter at high pressures [17]. Thus, the Hugoniot curve is the key to investigating the dynamic compression properties of materials.

Calculating the Hugoniot curve is an important application of the Grüneisen parameter [11,17]. According to the sodium chloride's Grüneisen parameters we got, the calculated Hugoniot curve of the NaCl B1 phase for the Hugoniot appears up to 20 GPa and 960 K. As Fig. 5 shows, the present pressures estimated for the fixed $V/V_0$ and $T$ values compare very well with the shock compression experiment data by Fritz et al. and other calculation works. Fig. 5 shows that the difference between our work and the shock compression value is relatively significant at low pressure. This difference might be partly due to the relatively large error in our Grüneisen parameter versus volume curve fitting at low pressures, illustrating the importance of accurate Grüneisen parameter measurement for determining the Hugoniot curve.

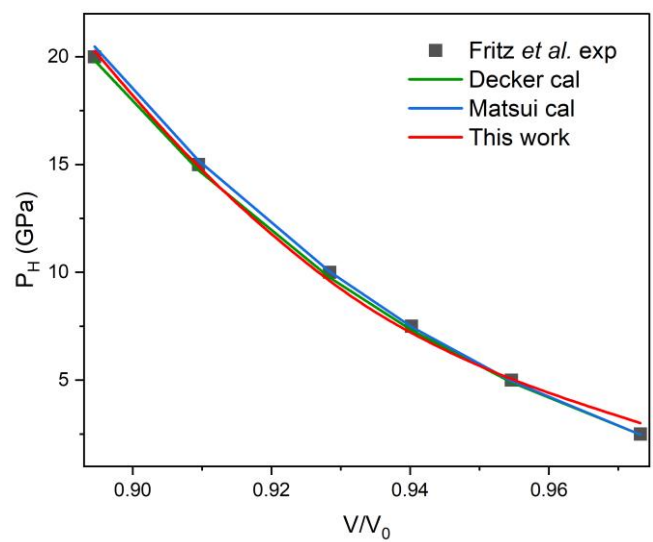

FIG. 5. The pressure along the Hugoniot curve of NaCl.

In this work, the Grüneisen parameter of B1-phase sodium chloride was measured within a wide pressure range (3.8-21.33 GPa) for the first time by our remote-controlled device, combining a high-speed pressure jump with ultra-fast temperature detection. The very quick pressure jump, and carefully measured and corrected temperatures gave us more accurate results. The obtained Grüneisen parameter is important for studying the thermodynamic properties of matter, such as the Hugoniot curve. Finally, according to our obtained sodium chloride Gruneisen parameters, we calculated the Hugoniot curve of B1-phase NaCl up to 20 GPa and 960 K, which compares very well with other works. Our results indicate that our improved Grüneisen parameter measurement method is reliable and feasible. Future development will further extend the experimental pressure range and improve data acquirement accuracy, significantly expanding research for many other substances in geophysics, among other fields.

**Corresponding author**

*Corresponding author: xiao.dong@nankai.edu.cn

†Corresponding author: leisu2050@iccas.ac.cn

‡Corresponding author: gqyang@iccas.ac.cn

**Notes**

The authors declare that they have no competing interests.

**Acknowledgements**

This work was supported by the National Science Foundation of China (Nos. 21627802, 51722209, 21273206, 92263101, 12174200), 2020-JCJQ project (GFJQ2126-007), Science Challenge Project (No. TZ2016001), Key Research Project of Higher Education (Nos. 15A140016 and 2010GGJS-110), and the National Key R&D Program of China (No. YS2018YFA070119).

# Supplementary Material for
# Expanding the Pressure Frontier in Grüneisen Parameter Measurement: Study of Sodium Chloride

Jun Kong[1,2], Kaiyuan Shi[2], Xingbang Dong[2], Xiao Dong[1,*], Xin Zhang[2],

Jiaqing Zhang[2], Lei Su[2,3,†] and Guoqiang Yang[3,‡]

[1]*Key Laboratory of Weak-Light Nonlinear Photonics and School of Physics, Nankai University, Tianjin 300071, China*

[2]*Center for High Pressure Science and Technology Advanced Research, Beijing 100093, China*

[3]*Key Laboratory of Photochemistry, Institute of Chemistry, University of Chinese Academy of Sciences, Chinese Academy of Sciences, Beijing 100190, China*

**Additional experiment methods details**

For the dDAC, three symmetrically positioned piezoelectric actuators (PZT ceramics, Nanomotions, and modal PAL40V25) were simultaneously elongated after charging, applying the load to a DAC (diamond anvil cell) through a screw. To ensure that the dDAC motion and data collection was synchronized, an arbitrary function generator concurrently provides the designed driving signals into the dDAC and triggers signals to the high-speed detectors. One spectrograph (Andor, SR500i) coupled with an intensified charge-coupled device (ICCD, Andor, DH334T), a digital multimeter (KEITHLEY DMM7510 7½ DIGITAL MULTIMETER), and a camera were prepared for the time-resolved diagnostic. The DMM7501 measurement capability for DC voltage is 10 nV to 1010 V and can achieve 1000000 readings/sec. Our detection voltage is in the order of ~mV, and we got the best data quality using 10000 readings/sec. The pressure was calibrated by the fluorescence of ruby balls placed inside the sample chamber [1].

**Theoretical framework**

The incomplete Mie-Grüneisen equation of state (EOS) is widely used in explosion mechanics and high-pressure physics to model solids, and it has the form:

$$P(V,E) = P_{ref}(V) + \frac{\gamma(V)}{V}\big(E - E_{ref}(V)\big) \tag{1}$$

The original typical form utilizes cold pressure $P_C$ and cold energy $E_C$ as reference curves.

We can also use $P_H(V)$ and $E_H(V)$ as a reference curve:

$$P(V,E) = P_H(V) + \frac{\gamma(V)}{V}\big(E - E_H(V)\big) \tag{2}$$

The subscript "$H$" denotes them in the Hugoniot state. For a shock locus through the initial state $(V_0, E_0)$ with pressure $P_0$ and temperature $T_0$, we have:

$$E_H(V) = E_0 + \frac{1}{2}(P_H(V) + P_0)(V - V_0) \tag{3}$$

where in our work, $P_0 = 0$ GPa.

Menikoff [2] has established the relation between $E, T$ based on $C_V = $ constant. The expression is:

$$T(V,E) = T_0\phi(V) + \big(E - E_{S_0}(V)\big)/C_V \tag{4}$$

where $\phi(V)$ is an integrating factor whose expression is:

$$\phi(V) = \exp\left[-\int_{V_0}^{V}\frac{\gamma(V')}{V'}dV'\right] \tag{5}$$

Furthermore, the temperature along the isentrope is as follows:

$$T_S = T_0 \exp\left[-\int_{V_0}^{V} \frac{\gamma(V')}{V'} dV'\right] \tag{6}$$

where $T_0$ is initial temperature along the isentrope, whose value in our work is 300 K.

Based on the above equations, we can get:

$$P_H = \frac{2C_V(T_S - T_H)}{V_H - V_0} \tag{7}$$

In the present study, we used the Vinet-Rydberg EOS to calculate the values of $V/V_0$ as a function of pressure $P$ at room temperature. We used a third-order polynomial to fit $\gamma(V/V_0)$. The equations for $P$ based on the Vinet-Rydberg EOS are given below:

$$P = 3K_0 x^{-2}(1-x)\exp[\eta(1-x)] \tag{8}$$

where $x = (V/V_0)^{1/3}$ and $\eta = (3/2)(K_0' - 1)$.

Table S1. Parameters for calculating $P_H$.

| Parameters | $\gamma_0$ [3] | $c_V$ [4] | $K_0$ [5] | $K_0'$ [5] |
|---|---|---|---|---|
| NaCl (300 K, 0 GPa) | 1.62 | 0.8168×10$^7$ erg/(g·K) | 24 GPa | 5.5 |

As is shown in Fig. 2c, the temperature rise in the sample started almost synchronously with the pressure jump and both reached their maxima in nearly the same time. It is also noticed that the temperature decreased observable after maximum while the pressure was held stable without visible drop. Therefore, a temperature correction is needed.

First, the temperature-increase data was fitted, and we can get:

$$T_1(t) = a_0 + a_1 t + a_2 t^2 + a_3 t^3 \tag{9}$$

Then, the temperature-decrease data was fitted,

$$T_2(t) = b_0 + b_1 t + b_2 t^2 + b_3 t^3 \tag{10}$$

Therefore, we can get the temperature-decrease rate with time after pressure jump,

$$\frac{dT_2(t)}{dt} = b_1 + 2b_2 t + 3b_3 t^2 \tag{11}$$

Then, we can get the temperature-decrease rate versus temperature after pressure jump,

$$\frac{dT_2(t)}{dt} = c_0 + c_1 T + c_2 T^2 + c_3 T^3 \tag{12}$$

Such temperature-decrease rate is coequally existent during the pressure jump although it is covered by a higher temperature rise, because the heat conductivity around the sample is coequal both during and after the pressure jump [7]. Therefore, the temperature-decrease rate during pressure jump can be obtained by

$$\frac{dT'(t)}{dt} = c_0 + c_1 T_1(t) + c_2 T_1(t)^2 + c_3 T_1(t)^3 \tag{13}$$

The temperature change caused by heat loss during the pressure jump can be obtained by integrating Eq. (13) with selecting a tiny time interval,

$$\Delta T' = \int_{t_1}^{t_2} c_0 + c_1 T_1(t) + c_2 T_1(t)^2 + c_3 T_1(t)^3 \, dt \tag{14}$$

Therefore, the actual temperature change during pressure jump at selected time interval can expressed by

$$\Delta T = \Delta T_1 + |\Delta T'| \tag{15}$$

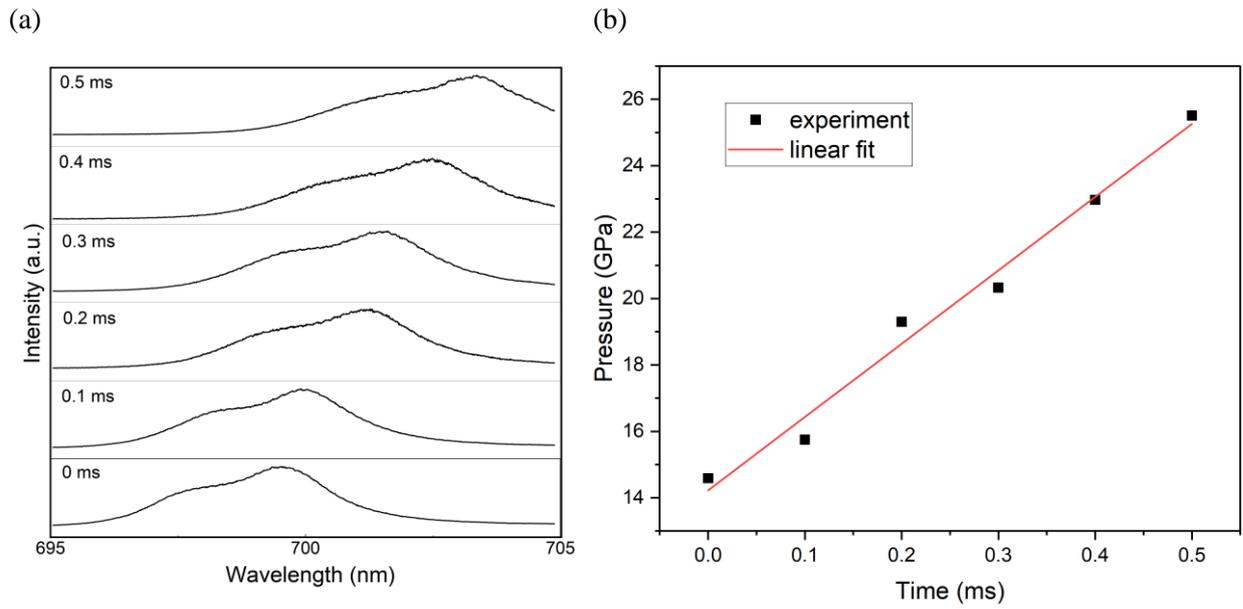

Figure S1. (a) The time-resolved fluorescence spectra of ruby collected under a fast compression. The exposure time is 80 µs, and the spectra were placed at an interval of 0.1 ms. (b) Pressure versus time. The change in pressure over time can be seen as a linear change.

Table S2. Experimental conditions and results for the Grüneisen parameter measurements of sodium chloride.

| Label | $P_m$ ($P_1$-$P_2$) (GPa) | $\Delta P$ (GPa) | $T(P_m)$ (K) | $\Delta T$ (K) | $K_s$ (GPa) | $\gamma$ | Error, $\Delta \gamma$ ($\pm$) |
|---|---|---|---|---|---|---|---|
| 1 | 7.24 (6.80-7.70) | 0.90 | 300.80 | 5.48 | 61.55 | 1.25 | 0.03 |
| 2 | 9.02 (8.84-9.20) | 0.36 | 300.66 | 1.97 | 70.45 | 1.26 | 0.03 |
| 3 | 7.63 (6.75-8.50) | 1.75 | 302.72 | 11.33 | 63.44 | 1.36 | 0.13 |
| 4 | 8.95 (7.39-10.50) | 3.11 | 302.07 | 17.27 | 70.09 | 1.29 | 0.02 |
| 5 | 11.25 (7.50-15.00) | 7.50 | 300.89 | 32.81 | 81.70 | 1.19 | 0.11 |
| 6 | 6.14 (5.30-6.97) | 1.67 | 302.89 | 12.30 | 55.93 | 1.36 | 0.08 |
| 7 | 5.13 (4.76-5.50) | 0.74 | 306.10 | 6.06 | 50.86 | 1.36 | 0.05 |
| 8 | 8.70 (5.30-12.10) | 6.80 | 305.52 | 41.07 | 68.85 | 1.36 | 0.05 |
| 9 | 6.87 (6.14-7.60) | 1.46 | 300.93 | 10.06 | 59.62 | 1.36 | 0.05 |
| 10 | 6.87 (6.30-7.44) | 1.14 | 300.93 | 7.64 | 59.62 | 1.33 | 0.06 |
| 11 | 6.77 (5.43-8.10) | 2.67 | 304.68 | 19.02 | 59.10 | 1.38 | 0.08 |
| 12 | 14.11 (7.56-20.65) | 13.09 | 313.28 | 50.51 | 96.09 | 1.18 | 0.06 |
| 13 | 20.00 (10.63-29.36) | 18.73 | 306.04 | 48.39 | 125.77 | 1.06 | 0.02 |
| 14 | 19.08 (8.56-29.60) | 21.04 | 299.73 | 59.58 | 121.15 | 1.14 | 0.04 |
| 15 | 17.37 (15.98-18.76) | 2.78 | 303.39 | 8.21 | 112.54 | 1.10 | 0.05 |
| 16 | 12.37 (10.34-14.40) | 4.06 | 299.59 | 16.96 | 87.34 | 1.22 | 0.07 |
| 17 | 20.63 (17.85-23.40) | 5.55 | 304.37 | 14.24 | 128.94 | 1.09 | 0.12 |
| 18 | 21.33 (18.98-23.67) | 4.69 | 304.55 | 10.99 | 132.47 | 1.02 | 0.11 |
| 19 | 3.80 (3.20-4.40) | 1.20 | 297.88 | 11.29 | 44.16 | 1.40 | 0.07 |
| 20 | 4.00 (3.40-4.60) | 1.20 | 300.65 | 10.56 | 45.17 | 1.32 | 0.06 |
| 21 | 11.39 (9.98-12.80) | 2.82 | 304.49 | 12.72 | 82.41 | 1.22 | 0.04 |
| 22 | 14.04 (12.97-15.10) | 2.13 | 302.63 | 7.93 | 95.73 | 1.18 | 0.07 |
| 23 | 15.18 (13.55-16.80) | 3.25 | 302.20 | 11.07 | 101.48 | 1.14 | 0.08 |
| 24 | 11.20 (10.50-11.90) | 1.40 | 299.80 | 6.40 | 81.45 | 1.24 | 0.04 |
| 25 | 13.65 (11.09-16.22) | 5.13 | 299.75 | 19.57 | 93.81 | 1.19 | 0.07 |

Note: $P_m$ is the mean pressure between the final ($P_2$) and initial ($P_1$) pressures. $\Delta P$ is the pressure jump; $T(P_m)$ is the instantaneous temperature corresponding to the median pressure; $\Delta T$ is the recorded temperature change after adding the temperature correction (see description in the main text and Supporting Information); and $K_s$ is the adiabatic bulk modulus [6].